\documentclass[12pt]{article}
\usepackage{amssymb}
\usepackage{epsfig}

\catcode`\@=11
\def\numberbysection{\@addtoreset{equation}{section}
        \def\theequation{\thesection.\arabic{equation}}}

\def\beq{\begin{equation}}
\def\eeq{\end{equation}}
\numberbysection
\begin{document}
\begin{titlepage}
\begin{center}
\hfill  \\
\vskip 1.in {\Large \bf Generating perfect fluid solutions in isotropic coordinates} \vskip 0.5in P. Valtancoli
\\[.2in]
{\em Dipartimento di Fisica, Polo Scientifico Universit\'a di Firenze \\
and INFN, Sezione di Firenze (Italy)\\
Via G. Sansone 1, 50019 Sesto Fiorentino, Italy}
\end{center}
\vskip .5in
\begin{abstract}
We explore the symmetry group of the pressure isotropy condition in isotropic coordinates finding a rich structure. We work out some specific examples.
\end{abstract}
\medskip
\end{titlepage}
\pagenumbering{arabic}
\section{Introduction}

There is an extensive literature describing perfect fluid solutions in general relativity. A perfect fluid is a fluid that is completely determined by its mass density
at rest $\mu$ and isotropic pressure $p$. Perfect fluids do not have shear stresses, viscosities or heat conduction.

The perfect fluids are used in general relativity as idealized models to describe the interior of a star such as white dwarfs and neutron stars, or as cosmological models (
examples are the dust ( $p = 0$ ) or the radiation fluid ( $\mu = 3 p$ )). A model of an isolated star generally consists of an internal region filled with fluid and an external region
which is an asymptotically flat vacuum solution. These two charts must be glued together through the outline of the star (the spherical surface of zero pressure).

For perfect fluid configurations Einstein's equations give rise only to the pressure isotropy condition. This condition together with the equation of state of the
fluid determines a regular distribution of matter in a unique way \cite{1}-\cite{2}. However this road turns out to be impractical for most of the equations of state and it is worthwhile to
follow other paths. One way is the transformation of the field equation into a form which is algebraic in one variable \cite{3}-\cite{4}, using the other variable as a generating function.
Kuchowicz \cite{5} does the same in isotropic coordinates, followed by a similar method due to Glass and Goldman \cite{6}-\cite{7}. They also gives sufficient conditions on the generating function which ensures that the resulting metric is interesting for physical applications.

Another possibility is introducing a method that can be used to generate new solutions from known ones \cite{8}-\cite{9}, which however requires calculations of integrals or solutions of
a differential equation ( see also more recently the series of papers \cite{10}-\cite{11}-\cite{12}-\cite{13} ).

In this article we focus on determining the symmetry group of the pressure isotropy condition elaborating from the method introduced in \cite{6}-\cite{7}, and we are able to find a sufficiently general algebraic structure that does not require the calculation of integrals or solve differential equations.

The article is organized as follows. First we introduce the general formalism and the new technique to generate perfect fluid solutions starting from those already known.
We discuss the physical conditions that the perfect fluid must satisfy in order that the solution is physically acceptable \cite{14}. We then move on to concrete applications.
We start from the famous Schwarzschild internal solution obtained through our technique starting from one elementary solution. Then our transformation is applied again to obtain a deformation of the Schwarzschild internal solution (already known in the literature) and we discuss in great detail the physical constraints. Finally in the last paragraph we introduce a completely new solution studying the space of the parameters in which it is physically acceptable.

\section{General formulas}

We want to describe here the relativistic stationary solutions of perfect fluid with the following choice of isotropic coordinates:

\beq ds^2 \ = \ - A^2(r) dt^2 + B^2(r) ( dr^2 + r^2 ( d\theta^2 + sin^2 \theta d \phi^2 )) \label{21}
\eeq

We define $ A^2(r) = e^{2 \nu (r)} $ and $ B^2(r) = e^{2 \omega (r)} $. Einstein's equations in the presence of a fluid perfect are defined by

\begin{eqnarray}
 8 \pi \mu & = & - e^{ -2 \omega } \left[ \ 2 \omega'' + {\omega'}^2 + \frac{4 \omega'}{r} \ \right] \nonumber \\
 8 \pi p & = & e^{ -2 \omega } \left[ \  {\omega'}^2 + 2 \omega' \ \nu' + \frac{2 \omega'}{r} + \ \frac{2 \nu'}{r} \ \right] \nonumber \\
 8 \pi p & = & e^{ -2 \omega } \left[ \  \omega'' + \nu'' + {\nu'}^2 + \frac{\omega'}{r} + \ \frac{\nu'}{r} \ \right]
\label{22} \end{eqnarray}

where $\mu$ and $p$ are respectively the density and the pressure of the perfect fluid. From these equations we derive the constraint due to the condition of pressure isotropy:

\beq \omega'' + \nu'' + {\nu'}^2 - 2 \omega' \nu' - {\omega'}^2 - \frac{\omega'}{r} - \ \frac{\nu'}{r} \ = \ 0
\label{23}
\eeq

In this article we want to discuss the symmetry group of this equation, which we will allow to generate new stationary perfect fluid solutions starting from known ones.

Define

\beq
\omega' \ = \ \frac{\alpha}{r} \ \ \ \ \ \ \ \ \ \nu' \ = \ \frac{\beta}{r}
\label{24}
\eeq

the pressure isotropy condition can be recast as

\beq
\beta^2 -\alpha^2 - 2 \alpha \beta - 2 ( \alpha + \beta ) + r ( \alpha' + \beta' ) \ = \ 0
\label{25}
\eeq

Define $ \widetilde{\alpha} \ = \ \alpha + \beta $, $ x = r^2 $, eq. (\ref{25}) can be rewritten as

\beq
2 \beta^2 \ = \ \widetilde{\alpha}^2 + 2 \widetilde{\alpha} - 2 x \frac{d}{dx} \ \widetilde{\alpha}
\label{26}
\eeq

Replace $ \widetilde{\alpha} = x / f(x) $, $ \beta =  x \ g(x) / f(x) $, the pressure isotropy condition becomes

\beq
\frac{d}{dx} \ f(x) \ = \ g^2 (x) - \frac{1}{2}
\label{27}
\eeq

Define

\beq
\widetilde{f} (x) \ = \ f(x) + \frac{x}{2}
\label{28}
\eeq

eq. (\ref{27}) reduces to

\beq
\frac{d}{dx} \ \widetilde{f} (x) \ = \ g^2 (x)
\label{29}
\eeq

It is now easy to study the symmetry group of the perfect fluid solutions with the choice of isotropic coordinates. This equation is in fact invariant with respect to the following
Moebius transformations:

\begin{eqnarray}
& \ & \widetilde{f}(x) \ \rightarrow  \ \frac{ a \widetilde{f}(x) + b }{ c \widetilde{f}(x) + d } \ \ \ \ \ \ \ \ \ a d \ - \ b c \ = \ 1 \nonumber \\
& \ & g ( x ) \ \rightarrow \ \frac{g(x)}{ c \widetilde{f}(x) + d }
\label{210}
\end{eqnarray}

We can thus build from a known solution a more extensive one ( depending on three arbitrary parameters ). Obviously many of these solutions will be discarded because they do not meet the necessary physical requirements (negative pressure and density, velocity of the sound greater than c, and so on ...) which we will discuss later. We note that under the Moebius transformations (\ref{210}) , the quantities $ \widetilde{\alpha} (x), \beta (x) $ transform in a much more complex way:

\begin{eqnarray}
& \ & \widetilde{\alpha}(x) \ \rightarrow  \ \frac{ ( \ 2 d x + c x^2 \ ) \ \widetilde{\alpha} + 2 c x^2 }{
(  \ a x + 2b - c x^2/2 - d x \ ) \ \widetilde{\alpha} + 2 a x - c x^2 } \ \ \ \ \ \ \ \ \ a d \ - \ b c \ = \ 1 \nonumber \\
& \ & \beta ( x ) \ \rightarrow \  \ \frac{ \ 2 x \beta (x) \ }{ ( \ a x + 2b - c x^2/2 - d x \ ) \ \widetilde{\alpha} + 2 a x - c x^2 }
\label{211}
\end{eqnarray}

Also the metric can be defined in terms of $ f $ and $ g $ :

\begin{eqnarray}
& \ &
\ln A^2 \ = \ \int \ dx \ \frac{ g(x )}{ f(x) }  \ + \ c_1
\nonumber \\
& \ &
\ln B^2 \ = \ \int \ dx \ \frac{ 1 - g(x) }{ f(x) }  \ + \ c_2
\label{212}
\end{eqnarray}

The other Einstein equations determine the pressure and density of the perfect fluid in terms of $ f(x) $ and $ g(x) $:

\begin{eqnarray}
p & = & \frac{1}{8 \pi f^2 B^2} \ [ \ 2 f - x ( g^2 -1 ) \ ]
\nonumber \\
\mu & = & 3 ( \ g-1 \ ) p - \frac{x g \ ( \ g^2-1 \ )}{8\pi f^2 B^2} + \frac{x g'(x)}{2\pi f B^2}
\nonumber \\
& = & ( \ 4g -3 \ ) \ p + \frac{1}{4\pi f B^2} \ ( \ 2 x g' - g \ )
\label{213}
\end{eqnarray}

which can be recast as

\beq
\mu \ = \ \frac{1}{8\pi f^2 B^2} \ [ \  6 f \ ( \ g - 1 \ ) - x \ ( \ g^2 -1 \ )( \ 4g-3 \ ) + 4x f g'  \ ]
\label{214}
\eeq

In the following we will need to know the derivatives of $ p $ and $ \mu $:

\beq
\frac{d p }{dx}  \ = \ \frac{1}{8\pi f^2 B^2} \ \left[ \ g \ ( \ 2 - 3g \ ) + \frac{x \ g}{f} \ ( \ 2g-1 \ )( \ g^2-1 \ ) - 2 x g g' \ \right]
\label{215}
\eeq

We can easily verify that the following relationship is satisfied:

\beq
\frac{d p }{dx}  \ = \ - \frac{g}{2f} ( p + \mu )
\label{216}
\eeq

Furthermore, the density derivative $ \mu $ can be calculated as follows:

\begin{eqnarray}
\frac{d \mu }{ dx } & = & \frac{1}{8\pi f^2 B^2} \ \left[ \ -5g \ ( \ 2g-1 \ )( \ g-1 \ ) -2 x g' ( \ 8 g^2 -5g -1 \ ) +  \right.
\nonumber \\
& + & \left. \frac{xg}{f} \ ( \ 2g-1 \ )( \ g^2-1 \ ) ( \ 4g-3 \ ) + 2f \ ( \ 5 g' + 2 x g'' \ ) \ \right]
\label{217}
\end{eqnarray}

Finally, the sound speed $ dp / d\mu $ can be represented as:

\beq
\frac{d \mu}{ dp } \ = \ 4 g - 3 \ + \ \frac{ g \ ( \ 2g^2-2g+1 \ ) - 2 xg' \ ( \ 4g^2 -2g -1 \ ) + 2f \ ( \ 5 g' + 2 x g'' \ )}{
- \ g \ ( \ 3g-2 \ ) +  ( x \ g / f  ) \ ( \ 2g-1 \ ) ( \ g^2-1 \ ) - 2x g g'}
\label{218}
\eeq

There are important physical constraints that the perfect fluid solutions must meet:

\begin{eqnarray}
& 1) & \infty > \mu > p > 0  \nonumber \\
& 2) &  p' < 0 \ \  \mu' <0 \nonumber \\
& 3) & \frac{ d p }{ d \mu } < 1 \nonumber \\
& 4) & \infty > A^2 > 0 \nonumber \\
& 5) & \infty > B^2 > 0
\label{219} \end{eqnarray}

Pressure and density must be positive and must be decreasing functions up to the radius where the pressure $ p $  becomes null ( outline of the star ).
The sound speed $ dp / d \mu $ must be less than the speed of light. The metrics $ A^2 $ and $ B^2 $ must be positive definite  and must not have singularities (even if apparent) as in the case of black hole. Note that a sufficient condition that satisfies all these physical requirements is that $ g'' > 0 $ ( see ref. \cite{6}-\cite{7} ).

Given generic $ f $ and $ g $ it is not always guaranteed that there is a surface at which $ p = 0 $

\beq 2 f \ = \ x_b \ ( \ g^2 -1 \ ) \ \ \ \ \ \ \ \longrightarrow \ \ \ \ \ x_b \ = \
\phi ( x_b ) = \frac{2 f}{g^2 -1}
\label{220} \eeq

For this to happen, condition $ \phi' \le 1 $ must be satisfied which in turn implies that

\beq g' \ge \frac{g \ ( \ g^2-1 \ )}{4 f}
\label{221} \eeq

The solution must then satisfy the condition that the sound speed is lower than light speed. In particular, it is important to discuss the following cases:

\begin{eqnarray} \frac{ d \mu }{ dp } \ ( 0 ) & > & 1 \ \ \ \ \ \ \ \rightarrow \ \ \ \ \ \ 0 < f_0 < \frac{g_0}{10 g_1} \ ( \ 10 g^2_0 -18 g_0 + 7 \ ) \nonumber \\
\frac{ d \mu }{ dp } \ ( x_b ) & > & 1
\label{222} \end{eqnarray}

where

\beq \frac{ d \mu }{ dp } \ ( x_b ) \ = \ \frac{6 g^2 -5 g + 1}{g}\ + \ \frac{ x \ ( \ g^2 -1 \ )}{g} \ \left( \frac{g' + 2 x g''}{g - 2 x g'}
\right)
\label{223} \eeq

\section{Schwarzschild interior solution}

This formalism also contains the black hole solution ( $ p = \mu = 0 $):

\begin{eqnarray}
\widetilde{f} ( x ) & = & \frac{2 x^2}{M^2} \nonumber \\
g(x) & = & \frac{2 \sqrt{x}}{M}
\label{31} \end{eqnarray}

We note that the black hole solution is not analytical in $ x = 0 $ and therefore even if we apply the Moebius transformation (\ref{210}), the perfect fluid solution generated by black hole is not physical (for example the pressure $ p $ can be negative ).

Now let's apply the general formalism of the previous section to get already known solutions. The most famous is undoubtedly Schwarzschild's internal solution. The latter one can be obtained from the following non-physical solution:

\begin{eqnarray}
\widetilde{f} ( x ) & = & x \nonumber \\
g(x) & = & 1
\label{32} \end{eqnarray}

applying the following Moebius transformation:

\begin{eqnarray}
A & = & \frac{3a-1}{2a} \nonumber \\
B & = & \frac{( \ a-1 \ )( \ 2a-1 \ )}{2ab} \ \ \ \ \ \ \ \ AD-BC = 1 \nonumber \\
C & = & - \frac{b}{a} \nonumber \\
D & = & \frac{1}{a}
\label{33} \end{eqnarray}

from which we obtain

\begin{eqnarray}
\widetilde{f} & \longrightarrow & \frac{ A \widetilde{f} + B }{ C \widetilde{f} + D } \ = \ \frac{ a^2 }{ b ( \ 1-bx \ )} \ + \ \frac{1-3a}{2b} \nonumber \\
g & \longrightarrow & \frac{g}{ C \widetilde{f} + D } \ = \ \frac{a}{ 1-bx }
\label{34} \end{eqnarray}

This is the internal solution of Schwarzschild in the isotropic coordinates. Let us recall some of its properties:

\beq f(x) \ = \ \frac{( 1 - 2a - bx )( 1-a -bx )}{2b ( 1-bx )} \ \ \ \ \ \ \ \ g(x) \ = \ \frac{ a }{ 1-bx } \label{35} \eeq

The metric takes the elementary form:

\beq A^2 \ = \ \left(  \frac{1-a-bx}{1-2a-bx} \right)^2 \ \ \ \ \ \ \ \ B^2 \ = \  \frac{1}{(1-2a-bx)^2} \label{36} \eeq

The condition $ \phi'(x) \leq 1 $ is always satisfied in any case. So there is an $ x = x_b $ for which the pressure is zero:

\beq x_b \ = \ \frac{2 f(x_b)}{g^2 (x_b) - 1} \ = \ \frac{ 2a-1 }{ b ( \ 3a-1 \ )} \label{37} \eeq

In general the pressure is:

\beq p \ = \ \frac{b}{2\pi} \ \frac{( \  1-2a - ( \ 1-3a \ ) bx \ ) }{ 1 - a - bx } \label{38} \eeq

and in the origin it is always greater than zero:

\beq p_0 \ = \ \frac{ 4b ( \ 1 - 2a  \ )}{( \ 1-a \ )} > 0 \label{39} \eeq

because $ a = 1 + \mu_0 / ( 3 p_0  ) \ > 1 $.

The density is constant:

\beq \mu = \mu_0 = 3 p_0 ( a-1 ) \label{310} \eeq

and therefore the solution is not physical because $ d \mu / dp \ = \ 0 $.

The connection with the external black hole solution for $ x = x_b$ implies that:

\beq g ( x_b ) \ = \ 3a -1 \ = \ \frac{2\sqrt{x_b}}{M} \ \ \ \ \ \ \ \longrightarrow \ \ \ \ \ \ \ \ \ x_b \ = \ \frac{M^2}{4} ( 3a-1 )^2 \label{311} \eeq

from which we derive that

\beq b \ = \ \frac{4}{M^2} \frac{ 2a-1 }{( 3a-1 )^3} \label{312} \eeq

The connection with the external solution fixes $ b = b(a) $ while the parameter $ a $ is linked to pressure and density at the origin.

\section{Extended Schwarzschild interior solution}

Let's start from Schwarzschild's internal solution

\beq \widetilde{f} \ = \ \frac{a^2}{b \ ( \ 1- bx \ )} \ + \ \frac{ 1-3a }{2b} \label{41} \eeq

and apply the following Moebius transformation:

\begin{eqnarray}
\widetilde{f} & \longrightarrow & \frac{  \widetilde{f} }{ 1 + k \widetilde{f} } \ = \ \frac{\widetilde{a}^2}{\widetilde{b} \ ( \ 1 - \widetilde{b}x \ )} \ + \ \frac{1+ 2 \alpha \ \widetilde{a}}{2 \widetilde{b}} \nonumber \\
g & \longrightarrow & \frac{ g }{ 1 + k \widetilde{f} } \ = \ \frac{\widetilde{a}}{( \ 1 - \widetilde{b} x \ )}
\label{42} \end{eqnarray}

a deformation of $ \alpha = - \ 3/2 $ where the old and new quantities are linked by the condition:

\begin{eqnarray}
\widetilde{a} & = & \frac{a}{D}  \nonumber \\
\widetilde{b} & = & \frac{b}{D} \left( \ 1 + k \left( \frac{ 1-3a }{ 2b } \right) \right) \nonumber \\
D & = & 1 + k \left( \frac{ 1-3a + 2 a^2 }{2b} \right) \nonumber \\
\alpha & = & -\frac{3}{2} - \frac{( \ D-1 \ )}{2a}
\label{43} \end{eqnarray}

This model has been already studied in ref. \cite{7}, but we will add some details. The perfect fluid solution that we will discuss in this chapter is therefore:

\begin{eqnarray}
f & = & \frac{a}{b} \ \left[ \ \frac{a}{z} + \frac{z}{2a} + \alpha \ \right] \ , \ \ \ \ \ \ \ \ z = 1-bx \nonumber \\
g & = & \frac{a}{z}
\label{44} \end{eqnarray}

The metric can be easily integrated leading to the following formulas:

\beq
A^2 \ = \ \left( \frac{ z + z_2}{ z + z_1} \right)^{-\frac{ 2a }{ ( z_1 - z_2 )} } \ \ \ \ \ \ \ \ \
B^2 \ = \ \frac{( z + z_2)^{ \frac{ 2(z_2+a)}{ (z_1-z_2)} } }{ ( z + z_1)^{ \frac{ 2(z_1+a)}{ (z_1-z_2)}}}
\label{45} \eeq

where $ z_{1,2} \ = \ a \ ( \ - \alpha \pm \sqrt{ \alpha^2 - 2} \ ) $.

Also this extended model has a value for which the pressure is zero. Indeed the condition $ 2f \ = \ x \ ( g^2-1 ) $ translates into the variable
 $ z $ as:

\beq  ( \ 1+ 2 \alpha a \ ) \ z^2  + 3 a^2 z - a^2 \ = \ 0 \label{46} \eeq

from which

\beq
z_b \ = \ \frac{ -3a^2 + a \sqrt{9 a^2 + 4( \ 1 + 2 \alpha a \ ))}}{2( \ 1 + 2 \alpha a \ )}
\label{47} \eeq

For $ \alpha = - \ 3/2 $ we obtain $ z_b = a / ( 3a-1 ) $.

It is easy to extrapolates the following particular values from the solution (\ref{44}):

\begin{eqnarray}
f_0 & = & \frac{1}{b} \ \left( \alpha a + a^2 + \frac{1}{2} \right) \nonumber \\
g_0 & = & a \ \ \ \ \ \ g_1 = ab
\label{48} \end{eqnarray}

from which the condition $ \frac{d \mu}{dp} (0) > 1 $ implies the following constraint on $ \alpha $ :

\beq - a - \frac{1}{2a} < \alpha < - 1.8 + \frac{0.2}{a} \label{49} \eeq

The sound speed, applying eq. (\ref{218}), becomes:

\beq \frac{d\mu}{dp} \ = \ \frac{4a}{z} \ - \ 3 \ + \ \frac{\frac{4a^2 ( \ 1+2\alpha +3a \ )}{z^3} + \frac{ 2a^2 ( \ \alpha -3 \ )+ 6a }{z^2}}{
\frac{2a}{z} - \frac{a^2}{z^2} - \frac{2a^2}{z^3} + \frac{2a ( \ 1-z \ )}{( \  z^2 + 2 \alpha a z + 2 a^2 \ )} \ \left( \ \frac{2a}{z} -1 \ \right) \ \left( \ \frac{a^2}{z^2} -1 \ \right)}
\label{410} \eeq

Obviously for $ \alpha = - \ 3 / 2 $ this expression simplifies and reduces to $ 0 $.

By plotting the solution (\ref{410}), we obtain that it satisfies the condition $ d \left( \ \frac{dp}{d\mu} \ \right) > 0 $ and therefore the condition $  \frac{dp}{d\mu} (x_b)   < 1  $
is more restrictive than the one at the origin.

Let us calculate  $ \frac{d\mu}{dp} ( x_b) $ at the point where the pressure becomes zero. The denominator of eq. (\ref{410}) simplifies considerably

\beq
\frac{d\mu}{dp} ( x_b) \ = \ \frac{4a}{z_b} \ - \ 3 \ + \ \frac{4a ( \ 1 + 2 \alpha + 3a \ ) \ + \ [ \  2a ( \ \alpha -3 \ ) + 6  \ ] z_b }{ a ( \ 3z_b - 2 \ )}  \ > \ 1
\label{411} \eeq

It is difficult to find a general formula for this condition and one has to resort to numerical calculation. However, for the following class of models an analytical constraint can be obtained:

\beq \alpha \ = \ - \ a \ + \ \frac{1}{2} \ \ \ \ \ \ \ \ \longrightarrow \ \ \ \ \ \ \ \ \ z_b \ = \ \frac{a}{1+2a} \label{412} \eeq

Now the constraint (\ref{411})  becomes calculable:

\beq
\frac{d\mu}{dp} ( x_b) \ > \ 1 \ \ \ \ \ \ \ \ \ \ \longrightarrow \ \ \ \ \ \ \ \ \ a \ > \ \frac{-1 + \sqrt{113}}{4} \ \simeq \ 2.4075
\label{413} \eeq

\section{New solution}

Finally, we introduce a new solution, inspired by equation (\ref{29}) and by the condition $ g'' > 0 $:

\begin{eqnarray}
\widetilde{f} ( x) & = & 2 k^2 \ e^{2x} \ \ \ \ \ \ \ \ \ \ k \ > \ 0 \nonumber \\
g(x) & = &  2k \ e^{x}
\label{51} \end{eqnarray}

Thus we have that $ f(x) \ = \ 2k^2 \ e^{2x} \ - \ x/2 $. This solution admits a zero of the pressure for

\beq x_b = 1 \label{52} \eeq

where the decreasing function $ \phi(x) $ satisfies the conditions:

\beq \phi_0 \ = \ \frac{4}{3} \ \ \ \ \ \ \ \ \ \phi_\infty \ = \ 1 \label{53} \eeq

The metrics $ A^2 $ and $ B^2 $ are not analytically calculable, but we can prove that $ B^2 $ is an regular increasing function of $ x $ in the range $ 0 < x < 1 $. The pressure can be calculated as follows:

\beq p \ = \ \frac{k^2}{ 2 \pi B^2 } \ \frac{( \ 1-x \ ) e^{2x}}{( \ 2 k^2 e^{2x} \ - \ \frac{x}{2} \ )^2} \label{54} \eeq

where $ p > 0 $ in the range $ 0 < x < 1 $ if $ k > 1 $.

We also calculate the derivative of the pressure with respect to the variable $ x $:

\beq
\frac{dp}{dx} \ = \ \frac{ k^2 e^{2x} }{ 2 \pi B^2 ( \ 2k^2 e^{2x} - \frac{x}{2} \ )^3 } \ \left[ \ 2 k^2 e^{2x} ( \ 2x - 3 \ ) + 2k e^{x} ( \ 1- x \ ) + x^2 - \frac{x}{2} \ \right]
\label{55} \eeq

The density can be easily calculated

\beq \mu \ = \ \frac{k}{2 \pi B^2 f^2} \ \left[ \ ( \ 8k e^{x} - 3 \ ) k e^{2x} ( \ 1-x \ ) + e^{x} ( \ 2x-1 \ ) ( \ 2k^2 e^{2x} - \frac{x}{2} \ ) \ \right]
\label{56} \eeq

Graphically we can check that it satisfies the condition $ \infty > \mu > p > 0 $ for $ k > 1 $.

The sound speed is in general

\begin{eqnarray} \frac{d \mu}{dp} & = & ( \ 8k e^{x} - 3 \ ) \ + \nonumber \\
& + & \frac{ 2k e^{x} ( \ 8 k^2 e^{2x} - 4k e^{x} + 1 \ ) - 4 k x e^{x} ( \ 16 k^2 e^{2x} - 4k e^{x} -1 \ ) + ( \ 4 k^2 e^{2x} - x \ )
( 10 k e^{x} + 4 k x e^{x} \ ) }{ -4k e^{x} ( \ 3k e^{x} - 1 \ ) + \frac{2 k x e^{x}}{( \ 2k^2 e^{2x} - \frac{x}{2} \ )} ( \ 4k e^{x}-1 \ ) ( \ 4 k^2 e^{2x} -1 \ ) - 8 k^2 x e^{2x} }
\nonumber \\
& \ & \label{57} \end{eqnarray}

We can analyze the sound speed at the origin:

\beq \frac{dp}{d\mu} (0) < 1 \ \ \ \ \ \ \ \ \ \ \longrightarrow \ \ \ \ \ \ \ \ k \ > \ \frac{9 + \sqrt{46}}{10} \simeq 1.578 \label{58} \eeq

Graphically this solution satisfies the condition

\beq d \left( \frac{dp}{d\mu} \right) < 0  \label{59} \eeq

and therefore the constraint $ \frac{dp}{d\mu} (x_b) < 1 $ does not give further constraints and it is always automatically verified if eq. (\ref{58}) is valid.

Finally we can add another parameter to the solution

\begin{eqnarray}
\widetilde{f} (x) & = & 2k^2 \ a \ e^{\frac{2x}{a}} \nonumber \\
g(x) & = & 2k \ e^{\frac{x}{a}} \label{510} \end{eqnarray}

by moving the zero of the pressure at point $ x = a $, but the solution is qualitatively the same as in case $ a = 1 $.

Let's apply the Moebius transformation (\ref{210}) on the solution (\ref{510}), we obtain

\begin{eqnarray}
\widetilde{f} (x) & = & k^2 \ a \ \tanh \left( \frac{x}{a} \right) \ + \ c  \nonumber \\
g(x) & = & \frac{ k }{\cosh (\frac{x}{a} )} \label{511} \end{eqnarray}

We immediately notice that $ g'' (x) < 0 $ around $ x = 0 $ ( the center of the star ), and graphing $ p(x) $ we see that it has a bizarre behavior and does not satisfy
to the condition $ \frac{dp}{dx} < 0 $, so let's classify the solution (\ref{511}) as not physical.

If we take (\ref{511}) as seed and apply the Moebius transformation (\ref{210})  we get again a three parameter extension of solution (\ref{511}), which may contain some physically
interesting ones.

\section{Conclusions}

Although most of perfect fluid solutions in general relativity have been studied in the Schwarzschild coordinates, we found a remarkable simplification
in the isotropic coordinates, where we have developed an algebraic method to generate new solutions starting from  already known ones. The symmetry group we have identified is
practically the group of Moebius transformations and we are therefore able to generate extended solutions with three arbitrary real parameters with respect to the original ones.
Obviously a strong selection of these solutions is necessary to meet the essential physical requirements ( positive pressure and density, sound velocity less than $c$ and so on ). We have
applied our method to the Schwarzschild interior solution and to a deformation of it, in which we have discussed the physical requirements in great detail. We think the method that
we have developed here could be useful in the future to identify new interesting solutions and to attempt a general classification of them.

\end{document}